# Comparative study of quantum methods in the resolution of track findings instances

Duy Dao DO[1], Hervé Kerivin[2], Philippe Lacomme[2], Bogdan Vulpescu[3]

[2] Université Clermont Auvergne,
1 rue de la Chebarde, 63177 Aubière, France
duy_dao.do@etu.uca.fr
[2] LIMOS - UMR CNRS 6158, Université Clermont Auvergne,
1 rue de la Chebarde, 63177 Aubière, France
herve.kerivin@uca.fr, phiilppe.lacomme@isima.fr
[3] Laboratoire de Physique de Clermont, Campus Universitaire des Cézeaux,
4 Avenue Blaise Pascal, 63178 Aubière, France
bogdan.vulpescu@clermont.in2p3.fr

**Abstract**. Track finding can be considered as a complex optimization problem initially introduced in particle physics involving the reconstruction of particle trajectories. A track is typically composed of several consecutive segments (track segments) that resembles a smooth curve without bifurcations. In this paper various modeling approaches are explored in order to assess both their impact and their effectiveness in solving them using quantum and classical methods. We present implementations of three classical models using CPLEX, two quantum models running on actual D-Wave quantum computers, and one quantum model on a D-Wave simulator. To facilitate a fair comparative study and encourage future research in this area, we introduce a new set of benchmark instances, categorized into small, medium, and large scales. Our evaluation of these methods on the benchmark instances indicates that D-Wave methods offer an excellent balance between computation time and result quality, outperforming CPLEX in numerous cases.

## 1 Introduction

A single collision between two protons from the two beams of the Large Hadron Collider (LHC) (Evans 2011) can generate thousands of new particles. The collision points can be identified with relatively precise coordinates using detectors with the innermost components built of multiple layers of silicon sensors arranged with cylindrical symmetry around the beam tube. This leads to a significant computational challenge in reconstructing the trajectories of charged particles starting from the small energy deposits in the detector, a process known as track finding.

With the High-Luminosity LHC (HL-LHC) (Apollinari 2015), the collision points are designed to generate an even more significant number of particles, further complicating the calculations due to the limited computational resources available with classical computers.

Quantum computing is an ambitious and rapidly evolving technology that offers the potential for exponential speedups. Current quantum devices, like D-Wave quantum computers, can now support up to several thousand qubits. Some authors claim that by harnessing the power of D-Wave quantum computers and their quantum annealing capabilities it could be possible to significantly reduce the computational time required to solve the track-finding problem (Rajak 2022).

In this paper, we propose a numerical experiment based on the Peterson's cost formula (Peterson 1989) with a Quadratic Constrained Binary Model (QCBM) and a Quadratic Unconstrained Binary Model (QUBM) solved using a quantum approach with D-Wave and a non-quantum approach with CPLEX, and a Binary Linear Program (BLP) solved only with CPLEX and a non-quantum approach. We have created a dataset extracted from the TrackML Kaggle challenge (Calafiura 2018), containing data simulated using ACTS (Ai 2019), and we have defined a set of instances composed of 10 small-scale instances with between 70 and 700 hits, of 10 medium-scale instances with 875 and 2450 hits and of 10 large-scale instances with 2625 and 4200 hits. The instances and the numerical experiments are available at the following web page

**https://perso.isima.fr/~lacomme/track_finding/data.html**

and are aimed to stimulate fair future research on the topic.

The paper is organized as follows. Section 2 details the construction of the QCBM model and its evolution into the QUBM and BLP models. Section 3 describes constructing the new instances based on key observations. Finally, Section 4 presents the results of applying our methods to these instances.

## 2    Modelizations

### 2.1    Problem description

Let us denote by $L = [p_1, p_2, \ldots p_L]$ the detector layers and by $HS = [h_1, h_2, \ldots, h_i, \ldots h_N]$ the ordered list of hits. A track is a set of consecutive segments, with a segment created by two hits on two different layers. Let us note that a track can be composed of hits of two non-consecutive layers since some high energy particle can miss leaving a trace in some layers. A track candidate is illustrated in Figure 1.

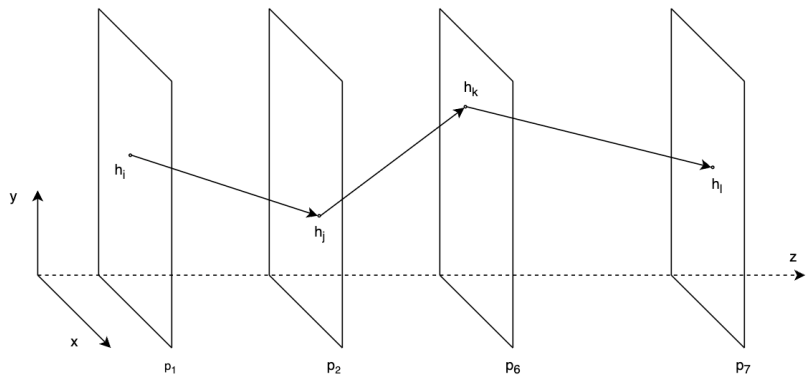

*Figure 1: An example of a track $t = [(h_i, h_j), (h_j, h_k), (h_k, h_l)]$*

As stressed in (Peterson 1989) two factors have to be taken into account to select segments for a track: the complementary angle created by two consecutive segments (we will call it later denoted $\beta$) and the total length of the two segments.

Due to the high velocity of the particles and the influence of the magnetic field, their trajectories will form either curves or straight lines. As a result, segments that are closer to the actual trajectory will have smaller $\beta$ angles. Therefore, the first objective is to evaluate the difference between the straight line and the track. Figure 2 illustrates two tracks, one in black and one in red, compared to the expected green trajectory. The red track has smaller $\beta$ angles than the black track, indicating that the red track is closer to the actual trajectory.

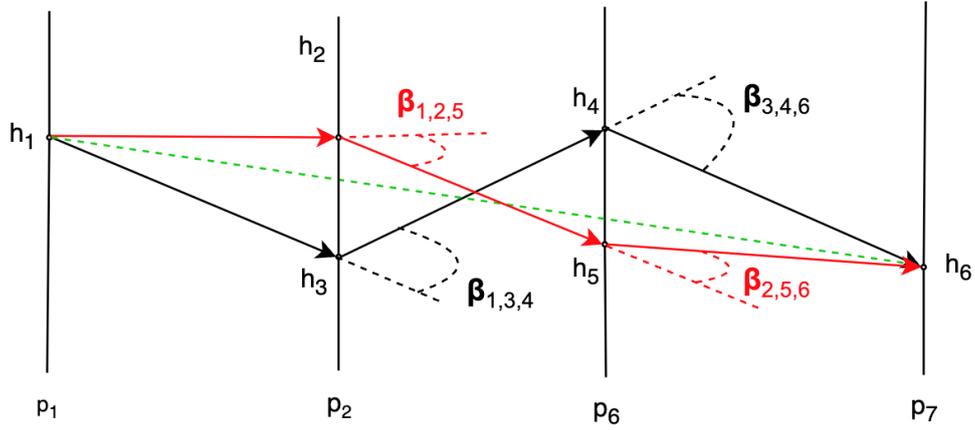

*Figure 2: The first criterion of a good track*

The second object is to choose between two tracks near the trajectory with the same sum of cos betas the cosines of the $\beta$ angles and the sum of the lengths of the segments is the deciding criterion. Consider the example in Figure 3 with two tracks, one in black with three adjacent segments and one in red with two adjacent segments. Let us suppose that $\cos(\beta_{1,2,5}) = \cos(\beta_{1,3,4}) + \cos(\beta_{3,4,5}) = \theta$, and $d_{1,2} = d_{1,3} = d_{3,4} = d_{4,5} = 100$, and $d_{2,5} = 200$. Then we have the costs associated with the black track and the red track as follows:

$$c_{red} = -\frac{\cos(\beta_{1,2,5})}{d_{1,2} + d_{2,5}} = \frac{-\theta}{300} > c_{black} = -\left(\frac{\cos(\beta_{1,3,4})}{d_{1,3} + d_{3,4}} + \frac{\cos(\beta_{3,4,5})}{d_{3,4} + d_{4,5}}\right) = \frac{-\theta}{200}$$

The cost associated with the black track is smaller than the cost associated with the red track, which indicates that a track with a larger number of small consecutive segments is favored.

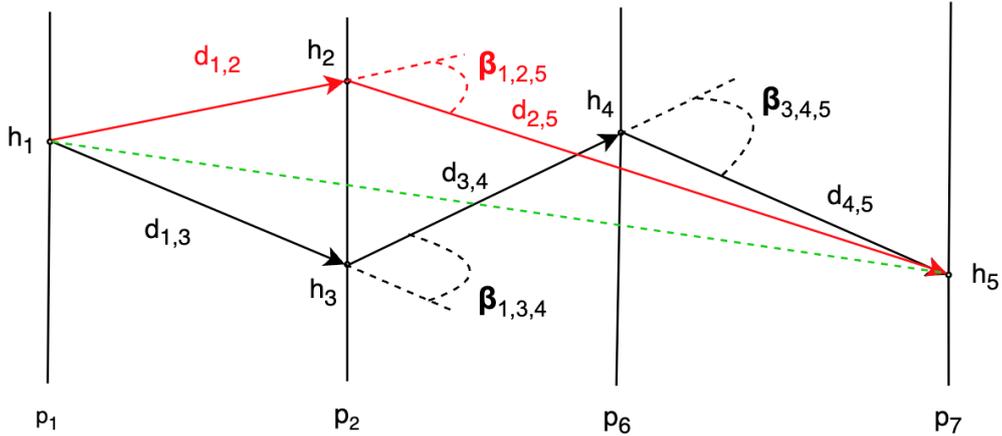

*Figure 3: The second criterion of a good track*

Let $c_{i,j,k}$ be the cost of two consecutive segments $\{(h_i, h_j), (h_j, h_k)\}$ is determined by the complementary angle $\beta_{i,j,k}$ and the lengths of the two segments $d_{i,j}, d_{j,k}$ describing in

$$c_{i,j,k} = \frac{-\cos(\beta_{i,j,k})}{d_{i,j} + d_{j,k}}. \quad (1)$$

Figure 4 illustrates an example of cost $c_{i,j,k}$.

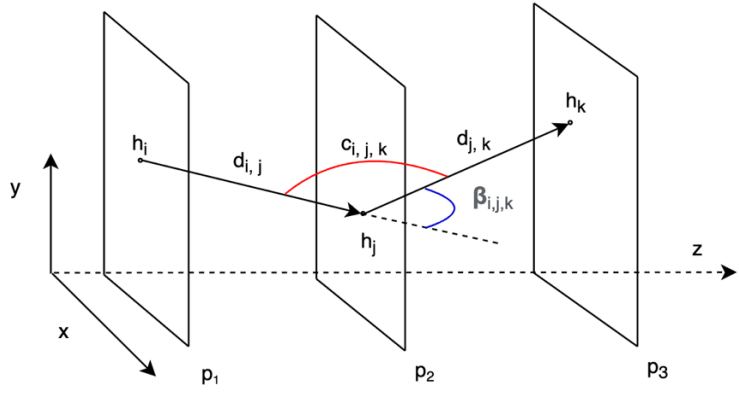

*Figure 4: An example of the cost of a track*

There are three reasons for choosing this formula of $c_{i,j,k}$: the first is if we assume a large length of the segments then $c_{i,j,k} \simeq 0$ ($\lim_{\substack{d_{i,j} \to \infty \\ d_{j,k} \to \infty}} c_{i,j,k} = 0$) and we converge to a solution that uses a large number of consecutive segments because the number of successive segments is more significant than when? the sum of distances will be is/gets? smaller; the second is that with smaller $\beta$ we have $\cos(\beta_{i,j,k}) \to 1$, so the track will get closer to a straight line, therefore we use $\cos(\beta_{i,j,k})$ to evaluate the difference between a track and a straight line; and the third is that by using a negative sign to the track cost, we can solve the problem of minimizing the cost of segments.

For example:

- Assuming the $\beta$ angles are the same: $\cos(\beta_{i,j,k}) = \cos(\beta_{i,j,l}) = \frac{\pi}{4}$ and the distances are different: $d_{i,j} = d_{j,k} = 10$ and $d_{j,l} = 100$. The costs are:

$$c_{i,j,k} = \frac{-\cos(\beta_{i,j,k})}{d_{i,j} + d_{j,k}} = \frac{-\frac{\pi}{4}}{10 + 10} = -\frac{\pi}{80} < c_{i,j,l} = \frac{-\cos(\beta_{i,j,l})}{d_{i,j} + d_{j,l}} = \frac{-\frac{\pi}{4}}{10 + 100} = -\frac{\pi}{440}$$

- Assuming the distances are similar $d_{i,j} = d_{j,k} = d_{j,l} = 10$ and the $\beta$ are different: $\cos(\beta_{i,j,k}) = \frac{\pi}{6}$ ; $\cos(\beta_{i,j,l}) = \frac{\pi}{4}$. The costs are:

$$c_{i,j,k} = \frac{-\cos(\beta_{i,j,k})}{d_{i,j} + d_{j,k}} = \frac{-\frac{\pi}{6}}{10 + 10} = -\frac{\pi}{120} < c_{i,j,l} = \frac{-\cos(\beta_{i,j,l})}{d_{i,j} + d_{j,l}} = \frac{-\frac{\pi}{4}}{10 + 10} = -\frac{\pi}{80}$$

**Notation used for the data (list indices start from 1)**

- $L$: the number of layers
- $HS$: an ordered list of hits
- $N$: the total number of hits
- $NL$: the number of hits in layers 1 to L-1 (excluding the last layer)
- $h_i$ : hit number $i$ in HS with $i = [1, N]$
- $\beta_{i,j,k}$: the angle between two segments $(h_i, h_j)$ and $(h_j, h_k)$.
- $d_{i,j}$: is the length of segment $(h_i, h_j)$
- $c_{i,j,k}$: the associated cost if the segments $(h_i, h_j)$ and $(h_j, h_k)$ are in the same track, with $c_{i,j,k} = -\frac{\cos(\beta_{i,j,k})}{d_{i,j}+d_{j,k}}$.

**The variables used for modeling the problem**

- $x_{i,j}$: a decision binary variable of for segment $(h_i, h_j)$ with

$$x_{i,j} = \begin{cases} 1, & \text{if the segment } (h_i, h_j) \text{ is assigned to a track} \\ 0, & \text{otherwise.} \end{cases}$$

## 2.2 Quadratic Constrained Binary Model – QCBM

We build the QCBM based on two parts: The objective function part and the constraint part.

The objective function part describes the objective which is to minimize the cost of all segments:

$$\boldsymbol{Min\ \left(\alpha \cdot \sum c_{i,j,k} \cdot x_{i,j} \cdot x_{j,k}\right)}, \quad \boldsymbol{\alpha > 0}, \quad (2)$$

The constraint part describes two conditions to build a track:

- For any hit $h_j$, we receive exactly one segment unit from hit $h_i$ :

$$\sum_i x_{i,j} = 1, \ \forall j$$

- For any hit $h_i$, we send exactly one segment unit to hit $h_j$ :

$$\sum_j x_{i,j} = 1, \ \forall i$$

- $x_{i,j}$ is binary $\forall i, j$.

This model can be solved by using CPLEX or as a Constrained Quadratic Model (CQM) with a D-Wave hybrid solver.

## 2.3 Quadratic Unconstrained Binary Model – QUBM

Based on the QCBM, we convert the constraints part to penalty parts using penalty terms and include it in the objective function to obtain the Hamiltonian of QUBM as follows:

**Constraints:**

- For any hit $h_j$, we receive exactly one segment unit from hit $h_i$ :

$$\sum_i x_{i,j} = 1, \ \forall j$$

- For any hit $h_i$, we send exactly one segment unit to hit $h_j$ :

$$\sum_j x_{i,j} = 1, \ \forall i$$

**Penalty parts:**

- Part 1:

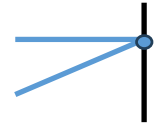

$$\gamma \cdot \sum_j \left(1 - \sum_i x_{i,j}\right)^2, \ \gamma > 0$$

- Part 2:

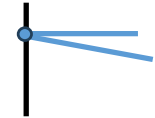

$$\gamma \cdot \sum_i \left(1 - \sum_j x_{i,j}\right)^2, \ \gamma > 0$$

**Hamiltonian:**

$$H = -\alpha \cdot \sum c_{i,j,k} \cdot x_{i,j} \cdot x_{j,k} + \gamma \cdot \sum_j \left(1 - \sum_i x_{i,j}\right)^2 + \gamma \cdot \sum_i \left(1 - \sum_j x_{i,j}\right)^2$$

$$\underbrace{\phantom{-\alpha \cdot \sum c_{i,j,k} \cdot x_{i,j} \cdot x_{j,k}}}_{\text{cost}} \quad \underbrace{\phantom{\gamma \cdot \sum_j \left(1 - \sum_i x_{i,j}\right)^2}}_{\text{penalty part 1}} \quad \underbrace{\phantom{\gamma \cdot \sum_i \left(1 - \sum_j x_{i,j}\right)^2}}_{\text{penalty part 2}}$$

Where: $\gamma, \alpha > 0$

This model can be solved by CPLEX or as a Binary Quadratic Model (BQM) with a D-Wave hybrid solver, and also using Simulated Annealing on a classical computer.

### 2.4  Binary Linear Program – BLP

Based on the QCBM we convert the objective function part to a linear form and add some constraints to the previous ones in order to obtain the BLP as follows:

Let us introduce new variables $z_{i,j,k} = x_{i,j} \cdot x_{j,k}$ which, together with the old ones, satisfy the constraints (Glover 1974):

- $x_{i,j} + x_{j,k} - z_{i,j,k} \leq 1$
- $z_{i,j,k} \leq x_{i,j}$
- $z_{i,j,k} \leq x_{j,k}$
- $0 \leq z_{i,j,k} \leq 1$

The objective is to minimize the cost of all segments:

$$Min \left(\alpha \cdot \sum c_{i,j,k} \cdot z_{i,j,k}\right), \; \alpha > 0.$$

with the constraints:

- For any hit $h_j$, we receive exactly one segment unit from hit $h_i$:
$$\sum_i x_{i,j} = 1, \;\; \forall j$$
- For any hit $h_i$, we send exactly one segment unit to hit $h_j$:
$$\sum_j x_{i,j} = 1, \;\; \forall i$$

- $x_{i,j} + x_{j,k} - z_{i,j,k} \leq 1$
- $z_{i,j,k} \leq x_{i,j}$
- $z_{i,j,k} \leq x_{j,k}$
- $0 \leq z_{i,j,k} \leq 1$
- $x_{i,j}$ is binary $\forall i, j$.

### 2.5  Models and solvers

Starting from our three models introduced above and the three available solvers we have used six methods as shown in Table 1.

Table 1: The allowed combinations of models and solvers, classical and quantum

| Models | Cplex | Dwave | |
|---|---|---|---|
| | Non Quantum resolution | Real Quantum Computer | Simulation of Quantum Computer |
| **Name of the method** | **MIQP** | **Hybrid solver** | **Simulated Annealing** |
| QUBM | Yes | Yes | Yes |
| QCBM | Yes | Yes | |
| BLP | Yes | | |

CPLEX can solve all three models. It allows us to solve optimization problems with only cost functions or cost functions and constraints. A hybrid solver[1] only allows us to solve problems of the form Binary Quadratic Model (BQM) without constraints, Constrained Quadratic Model (CQM), or Quadratic Model (QM) without constraints. This leads us to using BQM for QUBM and CQM for QCBM. Only BQM, Quadratic Unconstrained Binary Optimization (QUBO), or Ising models are accepted for simulated annealing. Thus, only QUBM can take advantage of this technology.

## 3  A new set of instances

A new set of instances is extracted from the *train_sample* dataset of the TrackML Particle Tracking Challenge[2]. Our goal is to study in the case where the detection layers have all the same number of hits.

We perform two steps to build this dataset:

- Extract hits from the original dataset;
- Preprocess step (build instances).

In the first step, we extract the hit data from the second sub-layer of each detector layer in volume number 9, collected from 10 events selected from a total of 100 events found in the dataset. From this, we isolate the hits associated with the same particle, resulting in a dataset where the number of hits per layer is consistent. Figure 5 illustrates the parameters for selecting hits, including the number of hits per layer, the percentage of hits chosen relative to the total hits per layer, and the minimum momentum.

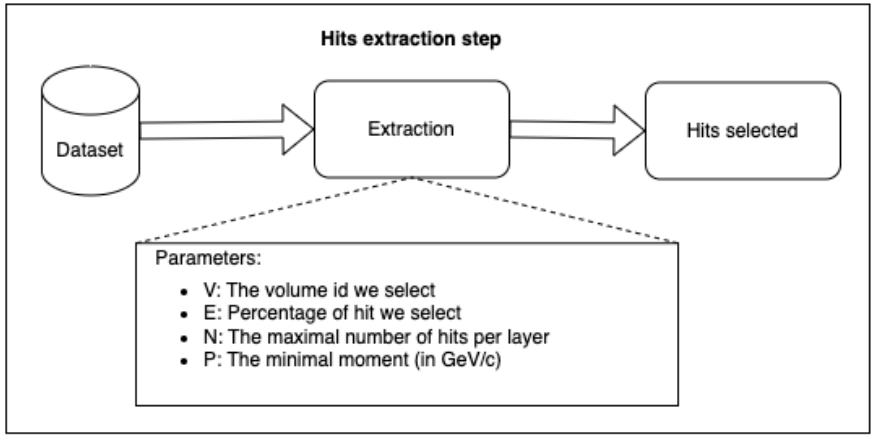

*Figure 5: The hits extraction step*

---

[1] https://docs.ocean.dwavesys.com/en/stable/concepts/index.html
[2] https://www.kaggle.com/c/trackml-particle-identification/

For example, V = 9, N = 10, E = 100 and P = 1 means that we extract all hits from volume 9, with a maximum of 10 hits per layer, a selection ratio of 100%, and a minimum acceptable momentum of 1. The momentum can be calculated using the following formula: $p_{min} \leq p = \sqrt{px^2 + py^2 + pz^2}$ with $p_{min}$ the minimum momentum and $px, py, pz$ the momentum components (in GeV/c) along each coordinate axis. In the second step we construct possible segments and pre-compute possible segment pairs. Segments are considered valid if their direction stays inside all the detection layers, as illustrated in Figure 6. We also limit the angle between two consecutive segments in order to reduce the number of generated segments from all $N!$ possible combinations of $N$ hits. Moreover, we impose that a good segment has two hits detected by two consecutive detector layers or with at most two missing layers in between.

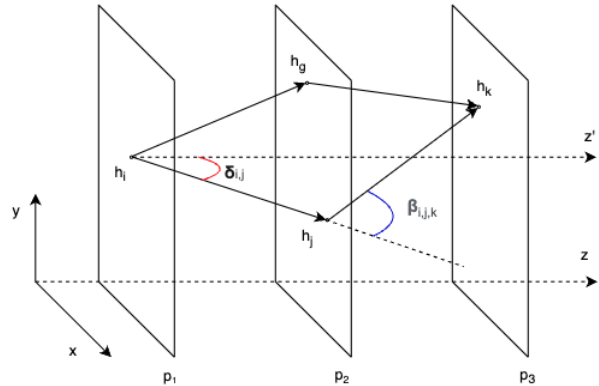

Figure 6: Build and filter segments

An instance is a set of tuples $(h_i, h_j, h_k)$ and $c_{i,j,k}$ corresponding costs. This study uses the datasets described in Table 2, with the preprocessing time shown in Figure 7. Details of each instance are described on our website[3].

Table 2: A new set of instances for track finding

| Datasets | No. instances | Range of no. tracks | Range of no. hits |
|---|---|---|---|
| Small-scale instances | 10 | 10-100 | 70-700 |
| Medium-scale instances | 10 | 125-350 | 875-2450 |
| Large-scale instances | 10 | 375-600 | 2625-4200 |

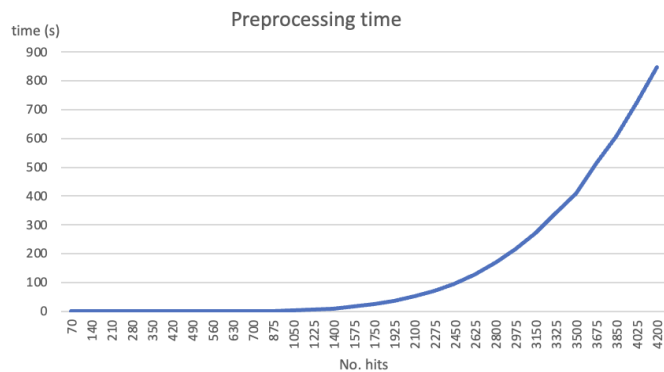

Figure 7: Preprocessing time of a new set of instances

The processing time for small datasets is almost negligible. From medium-scale instances onwards to large-scale instances, the time increases rapidly to $10^3$ seconds. We expect that with this preprocessing, the solving time will be small.

---

[3] https://perso.isima.fr/~lacomme/track_finding/data.html

# 4 Numerical experiments

## 4.1 Evaluation

This study evaluates the track finding performance based on the gap ratio between the true cost and the computed cost as follows:

$$gap = \frac{computed\ cost\ -\ true\ cost}{true\ cost}.100\%$$

where

- gap is the ratio error cost between the true cost and the computed cost,
- true cost is the cost of tracks from the true hits information, and
- computed cost is the cost of tracks the we compute by one of our methods.

## 4.2 Parameters

This study used the parameters introduced in table 3.

*Table 3: Parameters for methods*

| Methods | $\alpha$ | $\gamma$ |
|---|---|---|
| A_QUBM | 100 | 1 |
| C_BLP | 100 | - |
| C_QCBM | 100 | - |
| C_QUBM | 100 | 1 |
| D_QCBM | 100 | - |
| D_QUBM | 100 | 1 |

All methods use $\alpha = 100$ because $d_{i,j}$ is greater than or equal to 100 (in $\mu m$ units) and $\cos(\beta_{i,j,k}) \leq 1$, therefore the cost will have a small value. The QUBM model will be better without $\alpha$ ($\alpha = 1$), because with small $\alpha$ the cost values will form a fairly smooth landscape. This will make finding the optimal solution take longer. We use $\gamma = 1$ in the sense that the impact of the penalty functions is based on the number of segments themselves.

## 4.3 Results

This section will compare the results according to methods and datasets.

Denoted:

- S*: optimal solution cost
- S: solution cost found
- TP: preprocessing time (in seconds)
- TR: solution time (in seconds)
- TT: total time TT = TP + TR (in seconds)
- GAP: the error percentage: GAP = [(S - S*)/S*].100%
- ATT: average of total time
- AGAP: average of GAP
- PS: percentage of having solution followed by the method

*Table 4: Results of small-scale instances*

| | | Simulation of quantum computer | | | Non-quantum methods | | | | | | Quantum methods | | | |
|---|---|---|---|---|---|---|---|---|---|---|---|---|---|---|
| **LIBRARY USED** | | **D-Wave** | | | **CPLEX** | | | | | | **D-Wave** | | | |
| **Small-scale instances** | | **A_QUBM** | | | **C_BLP** | | **C_QCBM** | | **C_QUBM** | | **D_QCBM** | | **D_QUBM** | |
| **No_hits** | **S*** | **TP** | **TR** | **GAP** | **TR** | **GAP** | **TR** | **GAP** | **TR** | **GAP** | **TR** | **GAP** | **TR** | **GAP** |
| **70** | −17.35 | 0.00 | 0.08 | 0.00 | 0.01 | 0.00 | 0.00 | 0.00 | 0.01 | 0.00 | 0.03 | 0.00 | 3.06 | 0.00 |
| **140** | −34.70 | 0.00 | 0.22 | 0.00 | 0.00 | 0.00 | 0.01 | 0.00 | 0.01 | 0.00 | 0.03 | 0.00 | 3.08 | 0.00 |
| **210** | −52.05 | 0.01 | 0.15 | 0.00 | 0.00 | 0.00 | 0.01 | 0.00 | 0.01 | 0.00 | 0.04 | 0.00 | 3.13 | 0.00 |
| **280** | −69.39 | 0.02 | 0.19 | 0.00 | 0.05 | 0.00 | 0.01 | 0.00 | 0.03 | 0.00 | 5.11 | 0.00 | 3.20 | 0.00 |
| **350** | −86.73 | 0.04 | 0.23 | 0.00 | 0.01 | 0.00 | 0.01 | 0.00 | 0.03 | 0.00 | 5.11 | 0.00 | 3.17 | 0.00 |
| **420** | −104.09 | 0.07 | 0.29 | 0.00 | 0.01 | 0.00 | 0.01 | 0.00 | 0.03 | 0.00 | 5.11 | 0.00 | 3.20 | 0.00 |
| **490** | −121.45 | 0.15 | 0.38 | 0.00 | 0.06 | 0.00 | 0.05 | 0.00 | 0.10 | 0.00 | 5.12 | 0.00 | 3.39 | 0.00 |
| **560** | −138.80 | 0.24 | 0.47 | 0.00 | 0.07 | 0.00 | 0.06 | 0.00 | 0.13 | 0.00 | 5.35 | 0.00 | 3.29 | 0.00 |
| **630** | −156.16 | 0.38 | 0.73 | 0.32 | 0.10 | 0.00 | 0.09 | 0.00 | 0.21 | 0.00 | 5.31 | 0.00 | 3.71 | 0.00 |
| **700** | −173.50 | 0.57 | 1.29 | 0.43 | 0.14 | 0.00 | 0.10 | 0.00 | 0.31 | 0.00 | 5.20 | 0.00 | 3.51 | 0.00 |

*Table 5: Results of medium-scale instances*

| **Medium-scale instances** | | **A_QUBM** | | | **C_BLP** | | **C_QCBM** | | **C_QUBM** | | **D_QCBM** | | **D_QUBM** | |
|---|---|---|---|---|---|---|---|---|---|---|---|---|---|---|
| **No_hits** | **S*** | **TP** | **TR** | **GAP** | **TR** | **GAP** | **TR** | **GAP** | **TR** | **GAP** | **TR** | **GAP** | **TR** | **GAP** |
| **875** | −216.92 | 1.49 | 1.04 | 1.79 | 0.17 | 0.00 | 0.09 | 0.00 | 0.33 | 0.00 | 5.54 | 0.00 | 4.36 | 0.00 |
| **1050** | −260.29 | 2.95 | 1.54 | 3.22 | 0.43 | 0.00 | 0.21 | 0.00 | 1.61 | 0.00 | 5.50 | 0.00 | 6.11 | 0.29 |
| **1225** | −303.69 | 5.69 | 2.51 | 5.67 | 1.48 | 0.00 | 0.54 | 0.00 | 54.99 | 0.00 | 5.96 | 0.00 | 8.64 | 0.49 |
| **1400** | −347.06 | 9.50 | 3.35 | 9.34 | 5.88 | 0.00 | 2.06 | 0.00 | 93.09 | 0.00 | 5.63 | 0.00 | 11.12 | 1.40 |
| **1575** | −390.48 | 15.85 | 4.86 | 12.84 | 54.30 | 0.00 | 23.55 | 0.00 | 296.12 | 0.00 | 5.94 | 0.19 | 16.94 | 1.53 |
| **1750** | −433.87 | 24.22 | 6.87 | 17.41 | 128.78 | 0.00 | 73.35 | 0.00 | 360.04 | 0.00 | 6.33 | 0.22 | 25.05 | 2.40 |
| **1925** | −477.27 | 35.75 | 9.15 | 22.90 | 360.03 | 0.00 | 853.02 | 0.00 | 360.05 | 0.05 | 6.80 | 1.58 | 34.69 | 2.18 |
| **2100** | −520.68 | 52.24 | 13.45 | 31.10 | 360.06 | 0.00 | 360.06 | 0.00 | 360.09 | 8.07 | 7.29 | 5.54 | 48.31 | 4.91 |
| **2275** | −564.07 | 72.37 | 16.85 | 34.35 | 360.08 | 0.00 | 360.06 | 0.00 | 3600.13 | 12.22 | 8.09 | 13.07 | 70.33 | 3.33 |
| **2450** | −607.46 | 95.33 | 21.63 | 39.27 | 360.12 | 0.00 | 360.17 | 0.00 | 4503.13 | 31.72 | 9.30 | 38.79 | 89.72 | 3.80 |

*Table 6: Results of large-scale instances*

| Large-scale instances | | | A_QUBM | | C_BLP | | C_QCBM | | C_QUBM | | D_QCBM | | D_QUBM | |
|---|---|---|---|---|---|---|---|---|---|---|---|---|---|---|
| No_hits | S* | TP | TR | GAP | TR | GAP | TR | GAP | TR | GAP | TR | GAP | TR | GAP |
| **2625** | -650.86 | 128.87 | 27.92 | 42.00 | 3703.66 | 0.00 | 3600.23 | 0.00 | 3600.28 | 24.04 | 11.41 | 32.66 | 116.17 | 5.09 |
| **2800** | -694.26 | 169.28 | 44.76 | 45.34 | 3600.11 | 0.00 | 3600.07 | 0.00 | 4183.39 | 38.96 | 11.45 | 50.60 | 139.12 | 7.37 |
| **2975** | -737.63 | 215.67 | 52.43 | 47.14 | 4438.03 | 0.00 | 3600.07 | 0.00 | 3600.65 | 36.63 | 13.69 | 50.87 | 173.83 | 11.06 |
| **3150** | -781.04 | 270.89 | 82.35 | 50.99 | 3600.14 | 0.00 | 3600.25 | 0.00 | 3600.81 | 39.25 | 15.39 | 51.32 | 211.01 | 29.25 |
| **3325** | -824.43 | 338.92 | 104.86 | 51.95 | - | - | - | - | - | - | 17.86 | 54.87 | 271.68 | 15.05 |
| **3500** | -867.83 | 408.56 | 150.10 | 54.01 | - | - | - | - | - | - | 20.01 | 55.03 | 273.63 | 17.83 |
| **3675** | -911.24 | 513.66 | 243.59 | 56.16 | - | - | - | - | - | - | 24.11 | 56.62 | 303.63 | 17.83 |
| **3850** | -954.62 | 606.75 | 258.09 | 57.37 | - | - | - | - | - | - | 30.63 | 58.00 | 342.45 | 18.61 |
| **4025** | -998.00 | 723.46 | 353.11 | 58.63 | - | - | - | - | - | - | 33.07 | 58.89 | 387.00 | 21.58 |
| **4200** | -1041.38 | 848.31 | 340.60 | 59.62 | - | - | - | - | - | - | 31.84 | 78.99 | - | - |

*Table 7: Average results by methods*

| | | A_QUBM | | | C_BLP | | C_QCBM | | C_QUBM | | D_QCBM | | D_QUBM | |
|---|---|---|---|---|---|---|---|---|---|---|---|---|---|---|
| Dataset | ATP | ATT | AGAP | ATT | AGAP | ATT | AGAP | ATT | AGAP | ATT | AGAP | ATT | AGAP | |
| **Small-scale instances** | 0.15 | 0.49 | 0.10 | 0.19 | 0.00 | 0.18 | 0.00 | 0.23 | 0.00 | 3.72 | 0.00 | 3.14 | 0.00 | |
| **Medium-scale instances** | 31.54 | 39.67 | 17.79 | 194.67 | 0.00 | 234.85 | 0.00 | 994.50 | 5.21 | 36.66 | 5.94 | 58.35 | 2.03 | |
| **Large-scale instances** | 422.44 | 588.22 | 52.32 | 4150.69 | 2.02 | 3796.33 | 0.00 | 10362.06 | 150.18 | 432.45 | 54.78 | 537.75 | 15.96 | |

Table 4 shows that the solution time for non-quantum methods are is very short. In contrast, quantum methods take longer, but their processing times remain consistent and do not show significant increases. This indicates that, for a small number of hits, most methods deliver high accuracy and fast processing times.

In Table 5 the solution time for most methods increased sharply, except for D_QCBM and A_QUBM, which remained under 10 seconds. For C_BLP and C_QCBM, we limited the time that CPLEX could use for solving the problem to 360 seconds, and for C_QUBM the time was limited to 3600 seconds (excluding the time needed to create the Hamiltonian for CPLEX). While CPLEX delivers high accuracy, it comes at the cost of significantly increased solution time.

For A_QUBM and D_QCBM, the solution time increases more slowly, but the GAP (optimality gap) rises rapidly, particularly in the case of A_QUBM. With D_QUBM, although the solution time increases, it does so at a manageable rate, and the GAP remains relatively stable. Therefore, for medium-sized datasets, D_QUBM demonstrates the effectiveness of the quantum method in balancing processing time and GAP ratio.

In Table 6 many cells are empty for the following reasons. For methods using CPLEX, the resolution time is excessively long to achieve a small GAP, making it impractical to expect solutions within a reasonable processing time. Instead, we focus on analyzing A_QUBM and the two quantum methods. For A_QUBM, the resolution time does not increase rapidly, but the GAP remains consistently high. For quantum methods, there is a time limit on the D-Wave quantum computer for free accounts, restricted to 20 minutes per month. However, this is sufficient to solve the problem. With D_QCBM, the calculation time does not increase significantly because it leverages hybrid technology, handling constraints on traditional computers while solving the cost function on quantum computers. However, this comes at the expense of lower accuracy. On the other hand, D_QUBM, while taking longer to calculate, does so within a reasonable time frame and delivers relatively good accuracy.

Table 7 presents the average preprocessing time, average solution time, and average GAP across the datasets for different methods. The table indicates that CPLEX technology is the most effective for small datasets. However, leveraging D-Wave quantum computing technology is recommended for medium and large datasets.

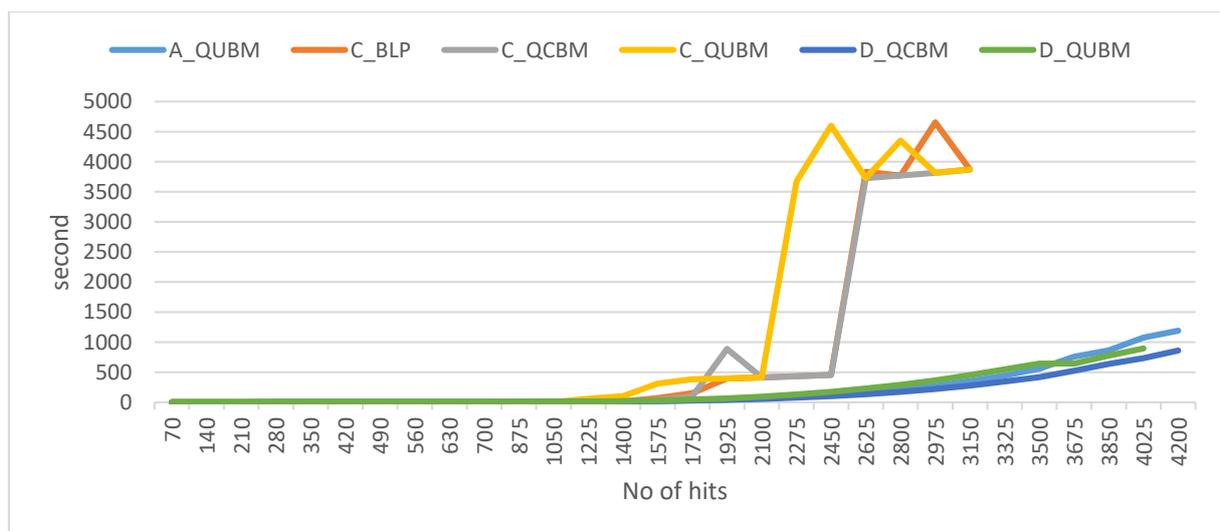

*Figure 8: Total time from preprocessing to resolution of six methods*

The total problem processing time is shown in Figure 8 where the time of non-quantum methods is very high while that of quantum methods is quite low.

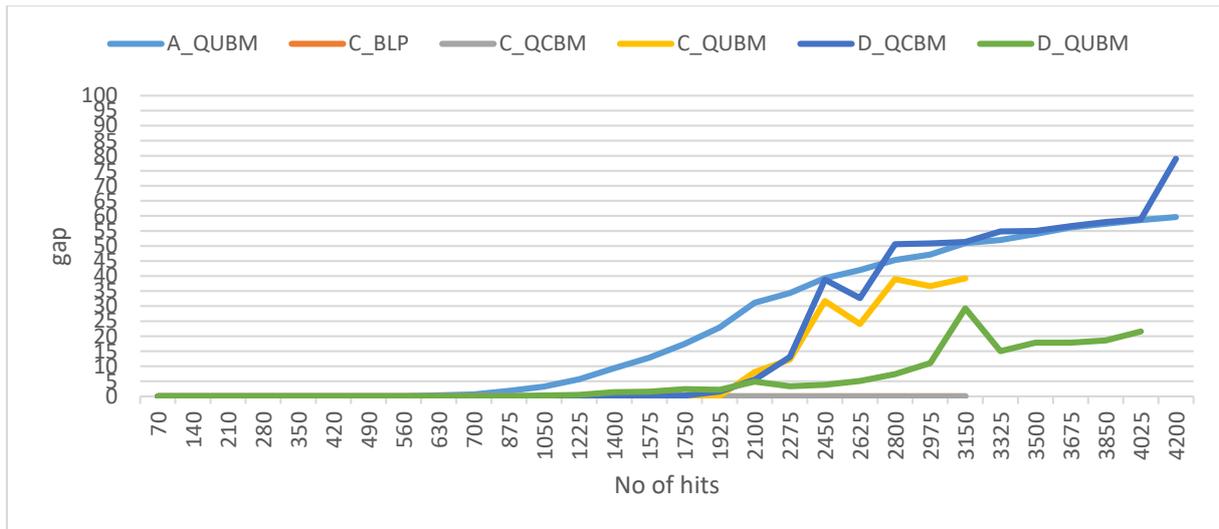

*Figure 9: Gap of the six methods*

Figure 9 shows the GAP ratio of six methods. In which, most of the methods using CPLEX have small gaps except for C_QUBM because this model has no constraints and only uses a penalty function. When the number of hits increases while the penalty term remains at 1, the value of the penalty function does not have much impact on the cost function. For A_QUBM, the GAP increases quite early and high. Meanwhile, for D_QUBM, the result is much better, the GAP is only about 22%. For D_QCBM, the initial result is not better than A_QUBM, but when the number of hits increases, the GAP also increases very quickly and is higher than A_QUBM. This can be seen that the QUBM model used on D-wave gives good results. However, the gamma parameter must also be adjusted when the number of hits increases.

## 5   Conclusion

In this study, we have developed and applied classical and quantum computing technologies to solve the track-finding problem using QCBM, QUBM, and BLP models. Our findings indicate that achieving high accuracy with non-quantum methods requires substantial solution time. Although Simulated Annealing on classical computers and the Hybrid method for solving CQM offer significantly shorter solution times, their accuracy is considerably lower. In contrast, solving the model using D_QUBM provides high accuracy within an acceptable solution time. The use of quantum computing will be a promising method to solve this problem in the future.